\documentclass[%
 reprint,
 amsmath,amssymb,
 aps,
]{revtex4-1}

\usepackage{graphicx}
\usepackage{dcolumn}
\usepackage{bm}

\begin{document}

\preprint{APS/123-QED}

\title{Holographic tool kit for optical communication beyond orbital angular momentum}

\author{Carmelo~Rosales-Guzm\'an}
\email{carmelo.rosalesguzman@wits.ac.za}
\affiliation{School of Physics, University of the Witwatersrand, Private Bag 3, Wits 2050, South Africa} 
\author{Abderrahmen~Trichili}
\affiliation{University of Carthage, Engineering School of Communication of Tunis (Sup'Com), GreS'Com Laboratory, Ghazala Technopark, 2083, Ariana,Tunisia} 
\author{Angela~Dudley}
\altaffiliation[Also at]{CSIR National Laser Centre, PO Box 395, Pretoria 0001, South Africa.}
\affiliation{School of Physics, University of the Witwatersrand, Private Bag 3, Wits 2050, South Africa}
 
\author{Bienvenu~Ndagano}
\affiliation{School of Physics, University of the Witwatersrand, Private Bag 3, Wits 2050, South Africa} 
\author{Amine~Ben~Salem}
\affiliation{University of Carthage, Engineering School of Communication of Tunis (Sup'Com), GreS'Com Laboratory, Ghazala Technopark, 2083, Ariana,Tunisia}

 \author{Mourad~Zghal}
  \altaffiliation[Also at ]{Institut Mines-T\'el\'ecom/T\'el\'ecom SudParis, 9 rue Charles Fourier, 91011 Evry, France}
 \affiliation{University of Carthage, Engineering School of Communication of Tunis (Sup'Com), GreS'Com Laboratory, Ghazala Technopark, 2083, Ariana,Tunisia}

\author{Andrew~Forbes}
 \affiliation{School of Physics, University of the Witwatersrand, Private Bag 3, Wits 2050, South Africa}

\date{\today}

\begin{abstract}
\noindent Mode division multiplexing (MDM) is mooted as a technology to address future bandwidth issues, and has been successfully demonstrated in free space using spatial modes with orbital angular momentum (OAM). To further increase the data transmission rate, more degrees of freedom are required to form a densely packed mode space.  Here we move beyond OAM and demonstrate multiplexing and demultiplexing using both the radial and azimuthal degrees of freedom.  We achieve this with a holographic approach that allows over 100 modes to be encoded on a single hologram, across a wide wavelength range, in a wavelength independent manner.  Our results offer a new tool that will prove useful in realising higher bit rates for next generation optical networks.\\
\end{abstract}

\maketitle


\section*{Introduction}

Since the beginning of the 21st century there has been a growing interest in increasing the capacity of telecommunication systems to eventually overcome our pending bandwidth crunch. Significant improvements in networks transmission capacity has been achieved through the use of polarization division multiplexing (PDM) and wavelength division multiplexing (WDM) techniques and also through implementing high order modulation formats \cite{WDM1, WDM2, WDM3}. However, it might not be possible to satisfy the exponential global capacity demand in the near future. One potential solution to eventually cope with bandwidth issues is space division multiplexing (SDM) \cite{Richardson1,Richardson2,Xia} and in particular the special case of mode division multiplexing (MDM), which was first suggested in the 1980s \cite{Berdague}. In MDM based communication systems, each spatial mode, from an orthogonal modal basis, can carry an independent data stream, thereby increasing the overall capacity by a factor equal to the number of modes used \cite{Li}. A particular mode basis for data communication is orbital angular momentum (OAM) \cite{Gibson} which has become the mode of choice in many studies due to its topical nature and ease of detection with phase-only optical elements \cite{Willner}.  Indeed, OAM multiplexing implementation results have reported Tbit/s transmission capacity over both free space and optical fibers \cite{Bozinovic,Wang1}. More recent reports have shown free space communication with a bit rate of 1.036 Pbit/s and a spectral efficiency of 112.6-bit/s/Hz using 26 OAM modes \cite{Wang2}. But, by taking into account the effects of atmospheric turbulence on the crosstalk and system bit error rate (BER) in an OAM multiplexed free space optics (FSO) link, experimental results have indicated that turbulence-induced signal fading will significantly deteriorate link performance and might cause link outage in the strong turbulence regime \cite{Turbulence1,Turbulence2,Turbulence3}. Recently, Zhao et al. claimed that OAM is outperformed by any conventional  mode division multiplexing technique with a complete basis or conventional line of sight (LOS) multipe-input multiple-output (MIMO) systems \cite{Zhao}. Indeed, OAM is only a subspace of the full space of Laguerre Gaussian (LG) beams where modes have two degrees of freedom: an azimuthal index $\ell$ and a radial index $p$, the former responsible for the OAM. In this study, we demonstrate a new holographic tool to realise a communication link using a densely packed LG mode set incorporating both radial and azimuthal degrees of freedom. We show that it is possible to multiplex/demultiplex over 100 spatial modes on a single hologram, written to a spatial light modulator, in a manner that is independent of wavelength. Our subset of the LG modes were successfully used as information carriers over a free space link to illustrate the robustness of our technique. The information is recovered by simultaneously detecting all 100 modes employing a single hologram.  Using this approach we are able to transmit several images with correlations higher than 98\%. Although our scheme is a proof-of-concept, it provides a useful basis for increasing the capacity of future optical communication systems. 

\section*{Results.}
\begin{figure*}[tb]
\centering
\includegraphics[width =.9\textwidth]{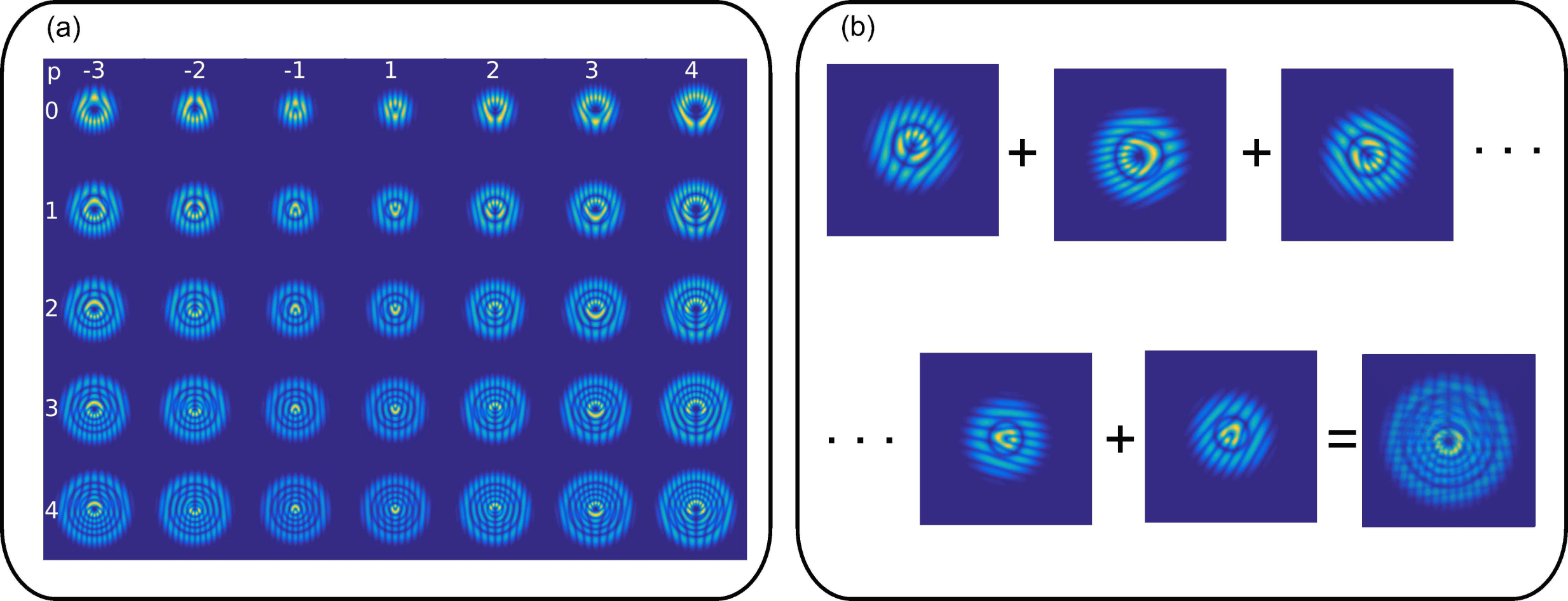}
\caption{{\bf Complex amplitude modulation and spatial-multiplexing.} (a) Holograms encoded via complex-amplitude modulation to generate different $\mathrm{LG}_{p\ell}$ modes. (b) Holograms encoded with different carrier frequencies are superimposed into a single hologram to produce a spatial separation of all modes in the Fourier plane.}
\label{holograms}
\end{figure*}

\begin{figure*}[tb]
	\centering
	\includegraphics[width = .9\textwidth]{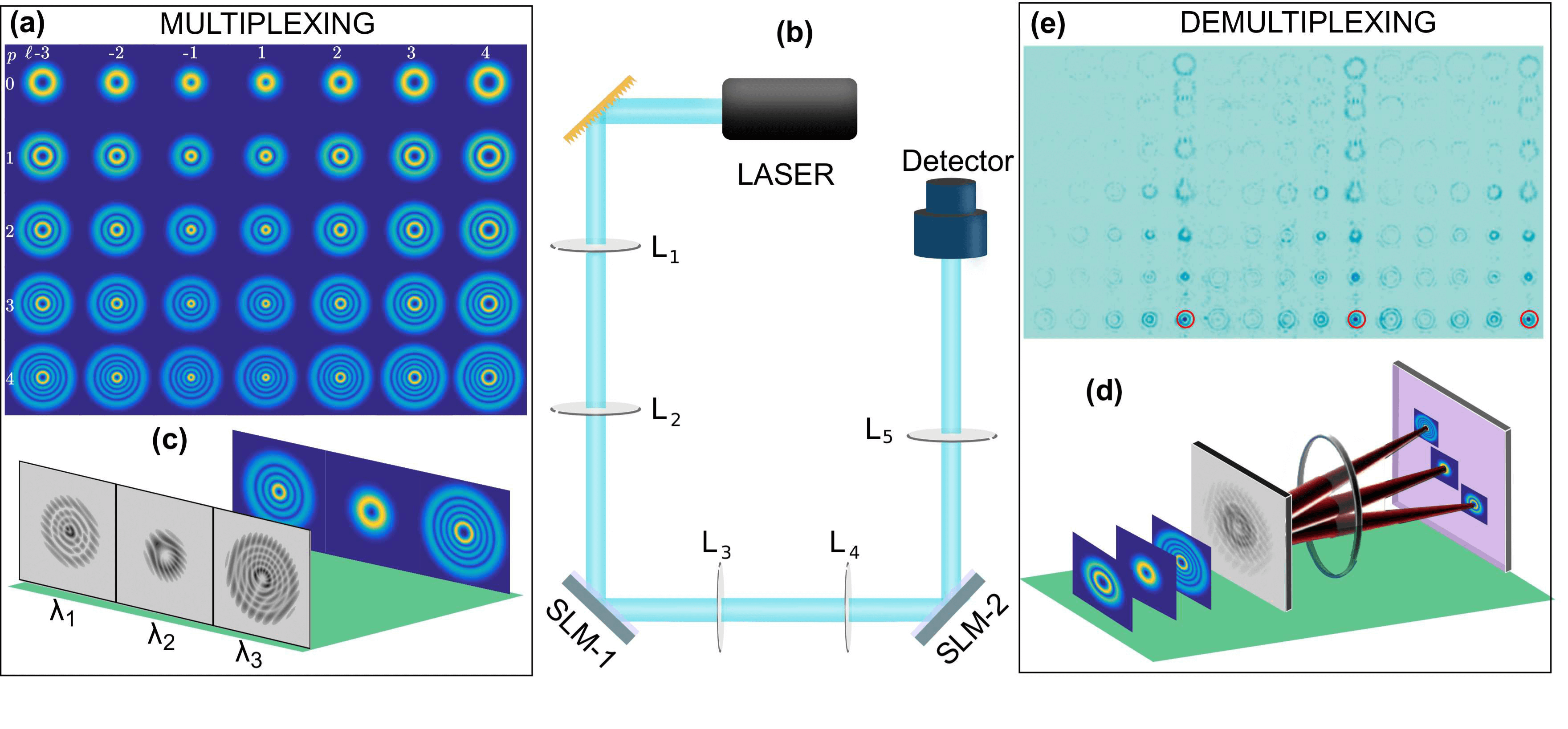}
\caption{{\bf Schematic of our Multiplexing and Demultiplexing setup.} (a) Intensity profiles of $\mathrm{LG}_{p\ell}$ modes generated from combinations of $p = \{0,1,2,3,4\}$ and $\ell = \{-3,-2,-1, 1,2,3,4\}$. (b) Experimental setup: Three components of a multiline Ion-Argon laser, $\lambda_1=457$ nm, $\lambda_2=488$ nm and $\lambda_3=514$ nm, are separated using a grating and sent to a Spatial Light Modulator (SLM-1). (c) The SLM is split into three independent screens, and addressed with holograms to produce the set of modes shown in (a). The information is propagated through free space and reconstructed in the second stage with a modal filter. (d) The modal filter consists of a superposition of all holograms encoded in SLM-2. (e) Each mode is identified in the far field using a CCD camera and a lens.}
\label{setup}
\end{figure*}

 Consider a LG mode in cylindrical coordinates, at its waist plane ($z=0$), described by:
\begin{eqnarray}
& &\mathrm{LG}_{p\ell}(r,\phi) =\sqrt{\frac{2p!}{\pi w_0^2(p+|\ell|)!}}\left(\frac{\sqrt{2}r}{w_0}\right)^{|\ell|}L_p^{|\ell|}\left(\frac{2r^2}{w_0^2}\right)\nonumber \\
 & &\times\exp\left(-\frac{r^2}{w_0^2}\right)\exp(i\ell \phi)
\label{eq:laguerre}
\end{eqnarray}

\noindent where $p$ and $\ell$ are the radial and azimuthal indices respectively, $(r,\phi)$ are the transverse coordinates, $L_p^{|\ell|}$ is the generalized Laguerre polynomial and $w_0$ is a scalar parameter corresponding to the Gaussian (fundamental mode) radius. The mode size is a function of the indices and is given by $w_{p\ell} = w_0 \sqrt{2p + |\ell| + 1}$. Such modes are shape invariant during propagation and are reduced to the special case of the Gaussian beam when $p=\ell =0$. This full set of modes can be experimentally generated using complex-amplitude modulation. For this experiment we use the CGH type 3 as described in \cite{Arrizon} to generate a subset of 35 $\mathrm{LG}_{p\ell}$ modes given by combination of $p = \{0,1,2,3,4\}$ and $\ell = \{-3,-2,-1, 1,2,3,4\}$. In this way, the amplitude and phase of the $\mathrm{LG}_{p\ell}$ modes set (Eq.\ref{eq:laguerre}) can be encoded into phase-only digital holograms and displayed on phase-only SLMs to generate any $\mathrm{LG}_{p\ell}$ mode. Moreover, the holograms can be multiplexed into a single hologram to generate multiple modes simultaneously. Figure \ref{holograms} (a) shows the generated holograms to create the desired subset of $\mathrm{LG}_{p\ell}$ modes for this experiment. Their corresponding theoretical intensity profile can be seen in  Fig. \ref{setup} (a).

The $\mathrm{LG}_{p\ell}$ modes generated in this way were encoded using three different wavelengths onto a single hologram, in a wavelength independent manner, and sent through free space. At the receiver, we were able to identify with high fidelity any of the 105 encoded modes in a single real time measurement, using a wavelength independent multimode correlation filter on a single SLM \cite{Flamm2013,Spangenberg}. This involves superimposing a series of single transmission functions $t_{n}(\textbf{r})$, each multiplied with a unique carrier frequency $\textbf{K}_{n}$ to produce a final transmission function $T(\textbf{r})$. 
\begin{eqnarray}
T(\textbf{r}) = \sum_{n=1}^{N}t_{n}(\textbf{r})\exp(i\textbf{K}_{n}\textbf{r})
\label{eq:multiplex}
\end{eqnarray}

\noindent where $N$ is the maximum number of multiplexed modes. In the Fourier plane the carrier frequencies $\textbf{K}_{n}$ manifest as separate spatial coordinates as illustrated in Fig. \ref{setup} (e). This approach allows multiple LG modes to be generated and detected simultaneously producing a high data transmission rate. The experimentally generated $\mathrm{LG}_{p\ell}$ modes are used to encode and decode information in our multiplexing and demultiplexing scheme as shown in Fig. \ref{setup}. 

\begin{figure}[h!]
	\centering
	\includegraphics[width=.4\textwidth]{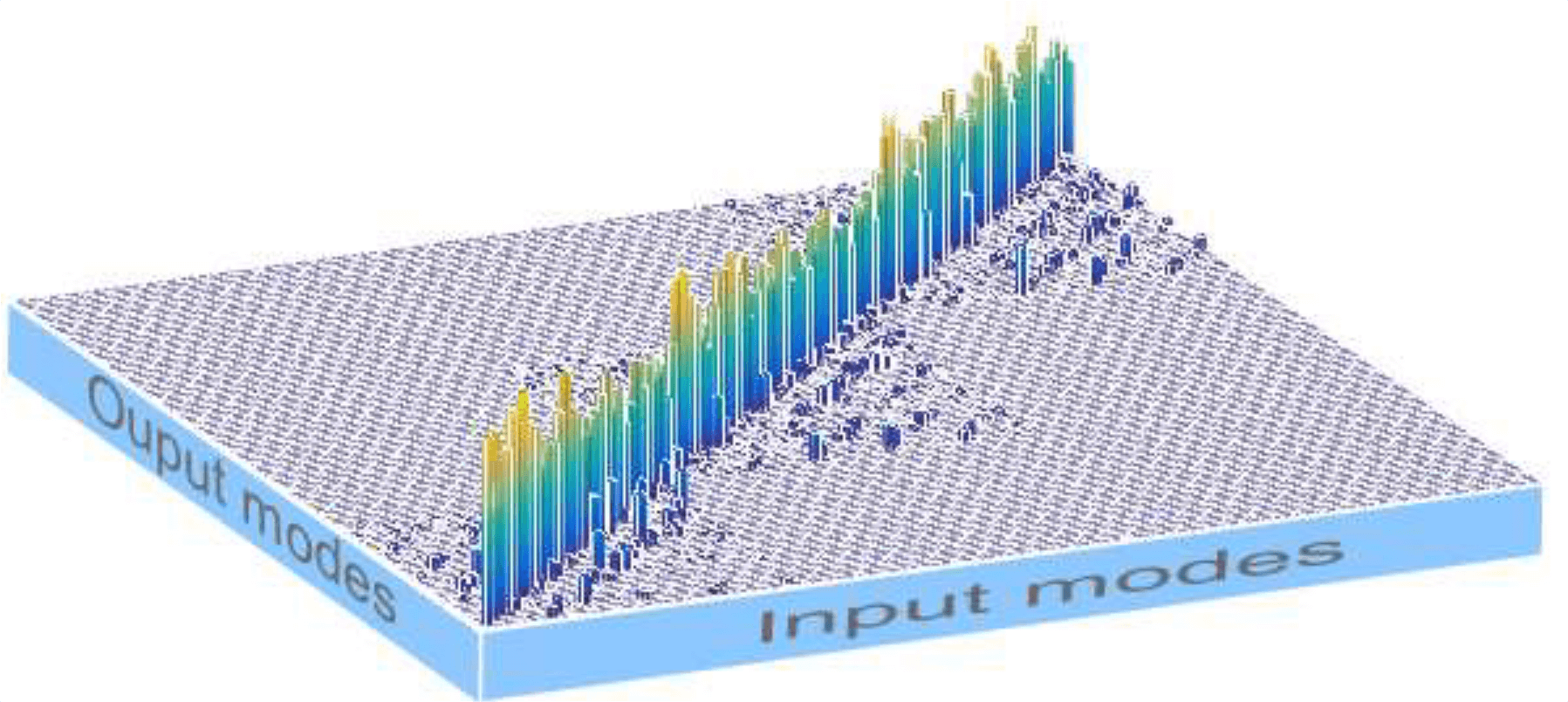}
	\caption{{\bf Cross Talk.} For each input mode we measure the output cross talk for all hundred and five output modes. In all cases the input mode is detected with very high accuracy, higher than 98\%.}
	\label{crosstalk}
\end{figure}

To date only the azimuthal component, responsible for the OAM content of these mode, has been used for data transmission, ostensibly because the divergence is lowest for $p=0$ \cite{Gibson,Zhao}. Here we demonstrate that the propagation dynamics, divergence being one example, is governed by the beam quality factor $M^2= 2p+|\ell|+ 1$ \cite{Forbes} and that modes with the same index will propagate in an identical manner regardless of the radial component $p$. For example the modes $\mathrm{LG}_{11}$ and $\mathrm{LG}_{04}$ will experience the same diffraction since both has the same value $M^2=4$. To show this, we encoded information in the set of $\mathrm{LG}_{p\ell}$ modes that incorporates both degrees of freedom, created as described before. Moreover, we multiplex the above mentioned subset of $\mathrm{LG}_{p\ell}$ modes on three different wavelengths to increase our (de) encoding basis set from 35 to 105.  All modes were generated using a single SLM (SLM-1 in Fig. \ref{setup} (b)) and a wide range multi-line laser. The data are encoded using these mode set and transferred in free space. This information is recovered by projecting the propagated information onto a modal filter. The modal filter consists of multiplexed holograms displayed on a second SLM (SLM-2)  and a CCD camera, capable of identifying with high accuracy any of the input modes (see experimental details). The intemodal crosstalk for the chosen modes, this is, the crosstalk between the input modes and the measured modes (output modes) is illustrated in Fig.~\ref{crosstalk}. As can be seen, the crosstalk between the different modes is very low and is independent of the $p$ value.

\begin{figure}[tb]
	\centering
	\includegraphics[width=.4\textwidth]{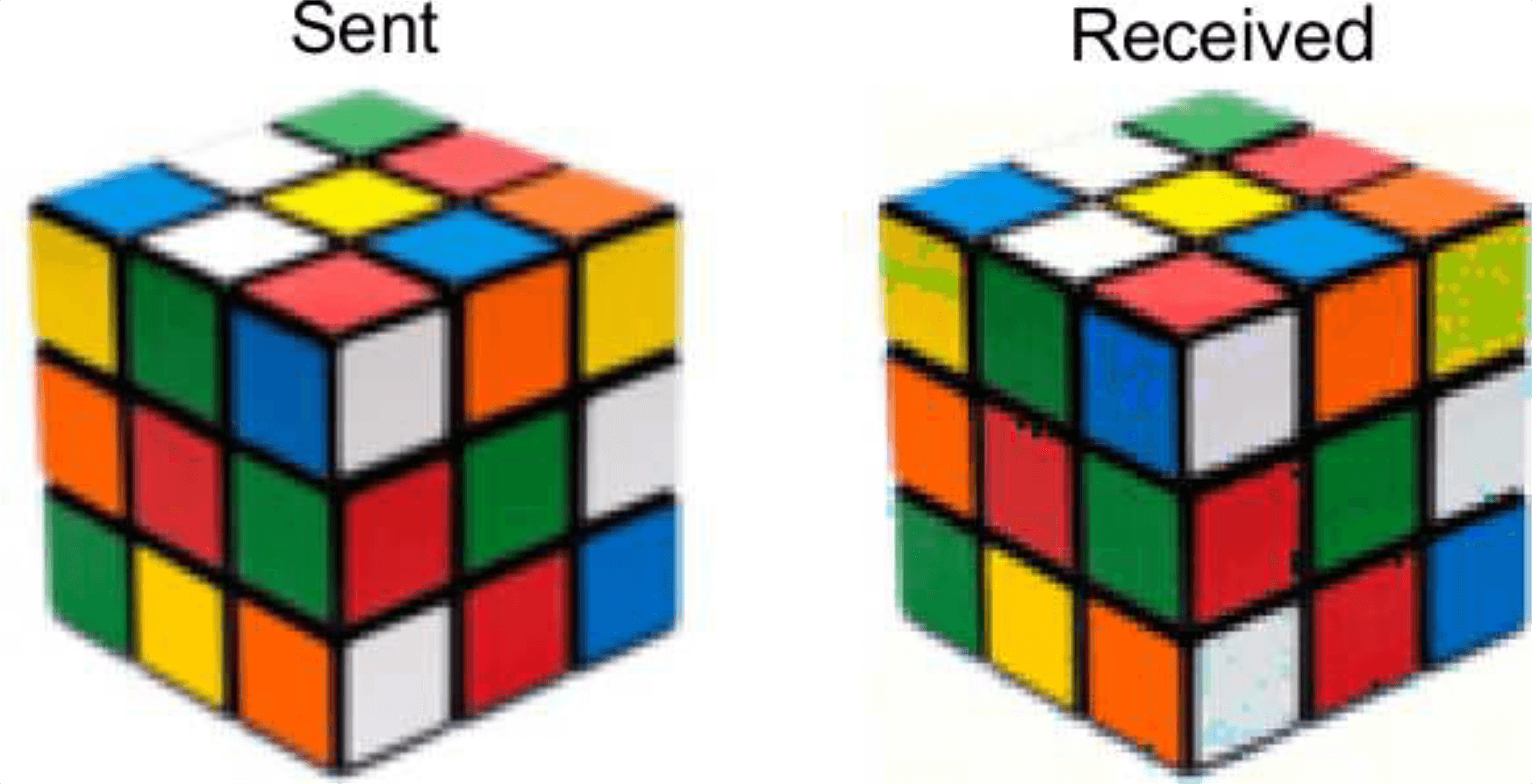}
	\caption{{\bf Example of sent and received images.} A quantification of the similarity between sent and received images is done using 2D image correlation. The value of the correlation coefficient ranges from 0 for nonidentical images to 1 for identical images. The correlation coefficient for the above image is $c=0.96$. Rubik's Cube$^\circledR$ used by permission of Rubik's Brand Ltd.}
	\label{cube}
\end{figure}

Figure \ref{cube} shows an example of an RGB image encoded, pixel by pixel as explained in the next section, and reconstructed in real time with a very high correlation coefficient ($c=0.96$). The correlation coefficient is a dimensionless number that measures the similarity between two images, being 0 for nonidentical images and 1 for identical images.

\subsection*{Encoding scheme}

\begin{figure*}[tb]
	\centering
	\includegraphics[width = 1\textwidth]{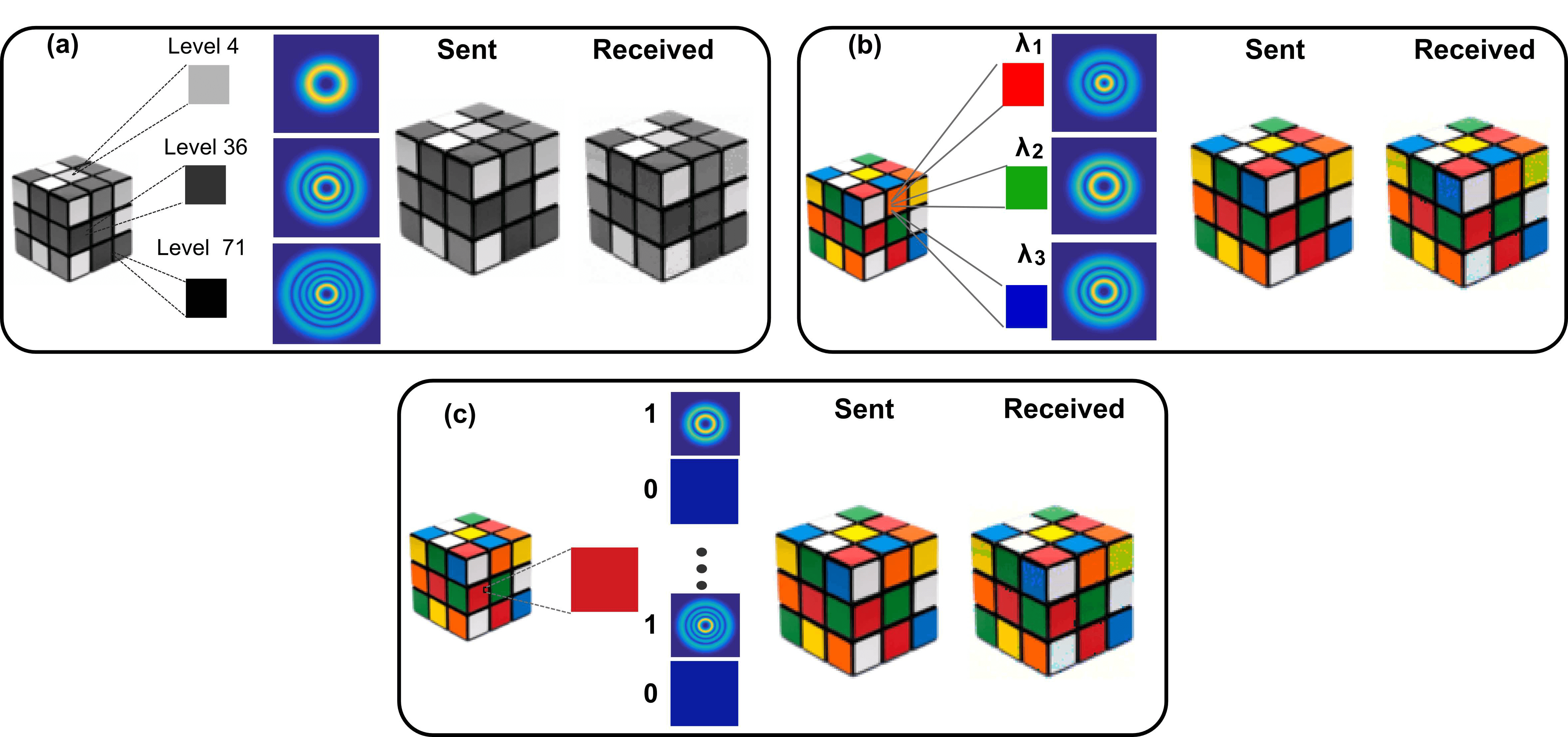}
	 \caption{{\bf Encoding Configurations.} (a) Single colour channel encoding, applied to gray-scale images. (b) RGB encoding, applied to colour images. (c)  Multi-bit encoding,  applied to both gray-scale and colour images. Rubik's Cube$^\circledR$ used by permission of Rubik's Brand Ltd }
	\label{coding}
\end{figure*}

The information encoding is performed in three different ways. In the first one, applied to grayscale images, we specifically assign a particular mode and a particular wavelength to the gray-level of each pixel forming the image. For example the mode $\mathrm{LG_{0-3}}$ generated with $\lambda_{1}$ is assigned to the lowest gray-level and the mode $\mathrm{LG_{44}}$ generated with $\lambda_{3}$ to the highest [see Fig.~\ref{coding} (a)]. In this approach we are able to reach 105 different levels of gray. In a second approach, applied to colour images, each pixel is first decomposed into its three colour components (red, blue and green). The level of saturation of each colour is assigned to one of the 35 different spatial modes and to a specific wavelength $\lambda_{1}$, $\lambda_{2}$ or $\lambda_{3}$ [see Fig. \ref{coding} (b)]. In this approach only 35 levels of saturation can be reached with a total number of 105 generated modes. Finally, in the third we implement multi-bit encoding [see Fig. \ref{coding} (c)].  In this scheme, 256 levels of contrast are achieved by multiplexing eight different modes on a single hologram. Each of the 256 possible  permutations, of these 8 modes, representing a particular gray level. Upon arrival to the detector each permutation is uniquely identified and the information is decoded to its 8-bit form to reconstruct the image. This approach was extended to high contrast colour images by using a particular wavelength for each primary colour intensity, achieving a total transmission rate of 24 bits per pixel. The transmission error rate, defined as the ratio between the number of wrong pixels and the total number of transmitted pixels, is found very low and did not reach 1\% in the case of gray-scale images. The reliability of our technique was further tested  by transmitting different complex images containing all levels of saturation in each RGB component. Here we only show the results for one image (Fig.~\ref{cube}), that clearly evinces the very high similitude between the original and recovered images.

\section*{Discussion.}

Very recently it was pointed out that OAM multiplexing is not an optimal technique for free-space information encoding and  that OAM itself does not increase the bandwidth of optical communication systems \cite{Zhao}. It has also been suggested that MDM, requires a complete mode set for a real bandwidth increment. Indeed, in all work to date only the azimuthal component of transverse modes, that gives rise to OAM, has been used in multiplexing schemes. Here we point out that the propagation dynamics (beam size, divergence, phase shift etc.) in free space are entirely governed by the beam quality factor, $M^2= 2p+|\ell|+ 1$ \cite{Forbes}, with analogous relations for fibre modes. The $M^2$ can be viewed as a mode index: modes with the same index (e.g., $p=0$, $\ell=2$ and $p=1$, $\ell=0$) will propagate in an identical manner as they have the same space-bandwidth product (see supplementary information for some examples). It is clear that one mode set will be as good as any other (at least in terms of perturbation-free communication), provided that the elements are orthogonal and regardless of whether it carries OAM or not. To demonstrate this, we create a mixed radial and azimuthal mode set from the $\mathrm{LG}_{p\ell}$ basis (with $p = \{0,1,2,3,4\}$ and $\ell = \{-3,-2,-1, 1,2,3,4\}$) and use this to transfer information over free space. Moreover, by implementing MDM on different wavelengths, we demonstrate that it is possible to expand the overall transmission capacity by several orders of magnitude. The number of carrier channels would be given by the number of optical modes times the number of wavelengths. In our experiment we generated 35 optical modes and combined this with 3 different wavelengths, creating a basis set of 105 modes. These modes are used as information carriers in a proof-of-concept free space link, capable of transmitting and recovering information in real time with very high fidelity.  Fig.~\ref{cube} is an example of the many images transmitted in our link. Each image is sent pixel by pixel, for this, the information of colour saturation of each pixel, is encoded using our mode set. Our encoding/decoding technique is key in the implementation of our optical link. Its simplicity linked to the versatility of SLMs, capable of operating in a wide range of the spectrum as well as with broad band sources, allowed us to generate customized digital hologram to encode and decode the information. Furthermore, the designed correlation filters are wavelength insensitive which allows the technique to operate in a large spectrum, compared to existing mode (de) multiplexers which are extremely wavelength sensitive, such as the photonic lantern. This approach can be extended to a wider range of wavelengths and to a higher number of modes. The use of polarization could be potentially an additional degree of freedom and could possibly double the overall transmission capacity of the system. Even though here we have used our modes as information carriers, this experiment establishes the basis for this technique to be incorporated into standard communication systems. In this case each mode would represent a channel that can be modulated and detected with conventional technology. 

To conclude, we have introduced a novel holographic technique that allows over 100 modes to be encoded/decoded on a single hologram, across a wide wavelength range, in a wavelength independent manner. This technique allowed us to incorporate the radial component of LG beams as another degree of freedom for mode division multiplexing. By combining both degrees of freedom, radial and azimuthal,  with wavelength-division multiplexing, we are able to generate over 100 information channels using a single hologram.  As a proof-of-concept, we implemented different encoding techniques to transmit information, with very high accuracy, in a free space link that employs conventional technology such as SLMs and CCD cameras. Our approach can be implemented in both, free space and optical fibres, facilitating studies towards high bit rate next generation networks.

\section*{Methods}
\subsection*{Experimental details.} 
The source, a continuum linearly-polarized Argon Ion laser (Laser Physics: 457-514 nm), is expanded and collimated by a telescope ($f_{1}=50$ mm and $f_{2}=300$ mm) to approximate a plane wave. Afterwards it is decomposed into its different wavelength components by means of a grating. Three of these components, $\lambda_{1}=457$ nm, $\lambda_{2}=488$ nm and $\lambda_{3}$=514 nm propagating almost parallel to each other, are redirected to a HoloEye Pluto Spatial Light Modulator (SLM, $1080\times1920$ pixels) with a resolution of 8 $\mathrm{\mu m}$ per pixel [see Fig.~\ref{setup} (b)]. The SLM is split into three independent screens, one for each beam, and controlled independently. Each third is addressed with a hologram representing a Laguerre-Gaussian mode ($\mathrm{LG}_{p\ell}$), where $p$ is the radial index and $\ell$ the azimuthal index [see Fig.~\ref{setup} (c)]. For this experiment we use 35 different modes  [see Fig.~\ref{setup} (a)], generated by combinations of $p$ = \{0, 1, 2, 3, 4\} and $\ell$ = \{-3, -2, -1, 1, 2, 3, 4\}. It should be stressed that the selection of the modes is made arbitrary and does not exclude any other combinations. These modes are encoded via complex amplitude modulations and only the first diffracted order from each third of the SLM is used. 

The information decoding is performed using modal decomposition, for this, the beams are projected onto a second SLM using a $4f$ configuration system ($f_{3}=150$ mm). This SLM is also split into three independent screens, each of which is addressed with a multiplexed hologram. This hologram consists of the complex conjugated of all 35 modes, encoded with different spatial carrier frequencies [see Fig.~\ref{setup} (d)]. To identify each mode, and therefore the graylevel of each pixel, we measured the on-axis intensity, of the projection, in the far field. For this we use a lens with focal length $f_4=200$ mm and a CCD camera (Point Grey Flea3 Mono USB3 $1280\times960$) in a $2f$ configuration system. In the detection plane (that of the camera), all 105 modes appear spatially separated, due to their unique carrier frequencies, in a rectangular configuration. In this way, an incoming mode can be unambiguously identified by detecting an on-axis high intensity [see Fig.~\ref{setup} (e)]. Even though, it is possible to get on-axis intensity for many other modes, the one that matches the incoming one, is always brighter. In our experiment, it is necessary to compensate for small spherical aberrations, this is done by digitally encoding a cylindrical lens on the second SLM which corrects for all modes. \newline


\begin{thebibliography}{10}
\expandafter\ifx\csname url\endcsname\relax
  \def\url#1{\texttt{#1}}\fi
\expandafter\ifx\csname urlprefix\endcsname\relax\def\urlprefix{URL }\fi
\providecommand{\bibinfo}[2]{#2}
\providecommand{\eprint}[2][]{\url{#2}}

\bibitem{WDM1}
\bibinfo{author}{Gnauck, A.~H.} \emph{et~al.}
\newblock \bibinfo{title}{Spectrally efficient long-haul {WDM} transmission
  using 224-{G}b/s polarization-multiplexed 16-{QAM}}.
\newblock \emph{\bibinfo{journal}{J. Lightwave Technol.}}
  \textbf{\bibinfo{volume}{29}}, \bibinfo{pages}{373--377}
  (\bibinfo{year}{2011}).

\bibitem{WDM2}
\bibinfo{author}{Zhou, X.} \emph{et~al.}
\newblock \bibinfo{title}{64-{T}b/s, 8 b/s/hz, {PDM}-36{QAM} transmission over
  320 km using both pre- and post-transmission digital signal processing}.
\newblock \emph{\bibinfo{journal}{J. Lightwave Technol.}}
  \textbf{\bibinfo{volume}{29}}, \bibinfo{pages}{571--577}
  (\bibinfo{year}{2011}).

\bibitem{WDM3}
\bibinfo{author}{Sano, A.} \emph{et~al.}
\newblock \bibinfo{title}{Ultra-high capacity {WDM} transmission using
  spectrally-efficient {PDM} 16-{QAM} modulation and {C}- and extended l-band
  wideband optical amplification}.
\newblock \emph{\bibinfo{journal}{J. Lightwave Technol.}}
  \textbf{\bibinfo{volume}{29}}, \bibinfo{pages}{578--586}
  (\bibinfo{year}{2011}).

\bibitem{Richardson1}
\bibinfo{author}{Richardson, D.~J.}
\newblock \bibinfo{title}{Filling the light pipe}.
\newblock \emph{\bibinfo{journal}{Science}} \textbf{\bibinfo{volume}{330}},
  \bibinfo{pages}{327--328} (\bibinfo{year}{2010}).

\bibitem{Richardson2}
\bibinfo{author}{Richardson, D.~J.}, \bibinfo{author}{Fini, J.~M.} \&
  \bibinfo{author}{Nelson, L.~E.}
\newblock \bibinfo{title}{Space-division multiplexing in optical fibres}.
\newblock \emph{\bibinfo{journal}{Nat Photon}} \textbf{\bibinfo{volume}{7}},
  \bibinfo{pages}{354--362} (\bibinfo{year}{2013}).
\newblock \bibinfo{note}{Review}.

\bibitem{Xia}
\bibinfo{author}{Li, G.}, \bibinfo{author}{Bai, N.}, \bibinfo{author}{Zhao, N.}
  \& \bibinfo{author}{Xia, C.}
\newblock \bibinfo{title}{Space-division multiplexing: the next frontier in
  optical communication}.
\newblock \emph{\bibinfo{journal}{Adv. Opt. Photon.}}
  \textbf{\bibinfo{volume}{6}}, \bibinfo{pages}{413--487}
  (\bibinfo{year}{2014}).

\bibitem{Berdague}
\bibinfo{author}{Berdagu\'{e}, S.} \& \bibinfo{author}{Facq, P.}
\newblock \bibinfo{title}{Mode division multiplexing in optical fibers}.
\newblock \emph{\bibinfo{journal}{Appl. Opt.}} \textbf{\bibinfo{volume}{21}},
  \bibinfo{pages}{1950--1955} (\bibinfo{year}{1982}).

\bibitem{Li}
\bibinfo{author}{Li, G.} \& \bibinfo{author}{Liu, X.}
\newblock \bibinfo{title}{Focus issue: Space multiplexed optical transmission}.
\newblock \emph{\bibinfo{journal}{Opt. Express}} \textbf{\bibinfo{volume}{19}},
  \bibinfo{pages}{16574--16575} (\bibinfo{year}{2011}).

\bibitem{Gibson}
\bibinfo{author}{Gibson, G.} \emph{et~al.}
\newblock \bibinfo{title}{Free-space information transfer using light beams
  carrying orbital angular momentum}.
\newblock \emph{\bibinfo{journal}{Opt. Express}} \textbf{\bibinfo{volume}{12}},
  \bibinfo{pages}{5448--5456} (\bibinfo{year}{2004}).

\bibitem{Willner}
\bibinfo{author}{Willner, A.~E.} \emph{et~al.}
\newblock \bibinfo{title}{Optical communications using orbital angular momentum
  beams}.
\newblock \emph{\bibinfo{journal}{Adv. Opt. Photon.}}
  \textbf{\bibinfo{volume}{7}}, \bibinfo{pages}{66--106}
  (\bibinfo{year}{2015}).

\bibitem{Bozinovic}
\bibinfo{author}{Bozinovic, N.} \emph{et~al.}
\newblock \bibinfo{title}{Terabit-scale orbital angular momentum mode division
  multiplexing in fibers}.
\newblock \emph{\bibinfo{journal}{Science}} \textbf{\bibinfo{volume}{340}},
  \bibinfo{pages}{1545--1548} (\bibinfo{year}{2013}).

\bibitem{Wang1}
\bibinfo{author}{Wang, J.} \emph{et~al.}
\newblock \bibinfo{title}{Terabit free-space data transmission employing
  orbital angular momentum multiplexing}.
\newblock \emph{\bibinfo{journal}{Nat Photon}} \textbf{\bibinfo{volume}{6}},
  \bibinfo{pages}{488--496} (\bibinfo{year}{2012}).

\bibitem{Wang2}
\bibinfo{author}{Wang, J.} \emph{et~al.}
\newblock \bibinfo{title}{{N}-dimentional multiplexing link with 1.036-{P}bit/s
  transmission capacity and 112.6-bit/s/{H}z spectral efficiency using
  {OFDM}-8{QAM} signals over 368 {WDM} pol-muxed 26 {OAM} modes}.
\newblock In \emph{\bibinfo{booktitle}{Optical Communication (ECOC), 2014
  European Conference on}}, \bibinfo{pages}{1--3} (\bibinfo{year}{2014}).

\bibitem{Turbulence1}
\bibinfo{author}{Ren, Y.} \emph{et~al.}
\newblock \bibinfo{title}{Atmospheric turbulence effects on the performance of
  a free space optical link employing orbital angular momentum multiplexing}.
\newblock \emph{\bibinfo{journal}{Opt. Lett.}} \textbf{\bibinfo{volume}{38}},
  \bibinfo{pages}{4062--4065} (\bibinfo{year}{2013}).

\bibitem{Turbulence2}
\bibinfo{author}{Xie, G.} \emph{et~al.}
\newblock \bibinfo{title}{Performance metrics and design considerations for a
  free-space optical orbital-angular-momentum-multiplexed communication link}.
\newblock \emph{\bibinfo{journal}{Optica}} \textbf{\bibinfo{volume}{2}},
  \bibinfo{pages}{357--365} (\bibinfo{year}{2015}).

\bibitem{Turbulence3}
\bibinfo{author}{Ren, Y.} \emph{et~al.}
\newblock \bibinfo{title}{Free-space optical communications using
  orbital-angular-momentum multiplexing combined with mimo-based spatial
  multiplexing}.
\newblock \emph{\bibinfo{journal}{Opt. Lett.}} \textbf{\bibinfo{volume}{40}},
  \bibinfo{pages}{4210--4213} (\bibinfo{year}{2015}).

\bibitem{Zhao}
\bibinfo{author}{Zhao, N.}, \bibinfo{author}{Li, X.}, \bibinfo{author}{Li, G.}
  \& \bibinfo{author}{Kahn, J.~M.}
\newblock \bibinfo{title}{Capacity limits of spatially multiplexed free-space
  communication}.
\newblock \emph{\bibinfo{journal}{Nat Photon}} \textbf{\bibinfo{volume}{9}},
  \bibinfo{pages}{822--826} (\bibinfo{year}{2015}).
\newblock \bibinfo{note}{Letter}.

\bibitem{Arrizon}
\bibinfo{author}{Arriz\'{o}n, V.}, \bibinfo{author}{Ruiz, U.},
  \bibinfo{author}{Carrada, R.} \& \bibinfo{author}{Gonz\'{a}lez, L.~A.}
\newblock \bibinfo{title}{Pixelated phase computer holograms for the accurate
  encoding of scalar complex fields}.
\newblock \emph{\bibinfo{journal}{J. Opt. Soc. Am. A}}
  \textbf{\bibinfo{volume}{24}}, \bibinfo{pages}{3500--3507}
  (\bibinfo{year}{2007}).

\bibitem{Flamm2013}
\bibinfo{author}{Flamm, D.} \emph{et~al.}
\newblock \bibinfo{title}{All-digital holographic tool for mode excitation and
  analysis in optical fibers}.
\newblock \emph{\bibinfo{journal}{J. Lightwave Technol.}}
  \textbf{\bibinfo{volume}{31}}, \bibinfo{pages}{1023--1032}
  (\bibinfo{year}{2013}).

\bibitem{Spangenberg}
\bibinfo{author}{Spangenberg, D.}, \bibinfo{author}{Dudley, A.},
  \bibinfo{author}{Neethling, P.~H.}, \bibinfo{author}{Rohwer, E.~G.} \&
  \bibinfo{author}{Forbes, A.}
\newblock \bibinfo{title}{White light wavefront control with a spatial light
  modulator.}
\newblock \emph{\bibinfo{journal}{Opt. Express}} \textbf{\bibinfo{volume}{22}},
  \bibinfo{pages}{13870--13879} (\bibinfo{year}{2014}).

\bibitem{Forbes}
\bibinfo{author}{Forbes, A.}
\newblock \emph{\bibinfo{title}{Laser beam propagation: generation and
  propagation of customized light}} (\bibinfo{publisher}{CRC Press},
  \bibinfo{year}{2014}).

\end{thebibliography}

\section*{Acknowledgments}
This work has been partially supported by Tunisian-South African bilateral project ``Towards spatial mode control in fibers for high bit rate optical communication,'' funded by the African Laser Centre (ALC). AT acknowledges financial support from the ICTP affiliated centre ``The Optical Society of Tunisia''. CRG acknowledges Claude Leon Foundation and CONACyT for funding.

\section*{Author contributions statement}
AF conceived the idea and supervised the project;  CRG, AD and AT conducted the experiment(s).  CRG prepared figures 1, 3, 4 and S5, CRG and BN prepared figures 2 and 5, AD prepared figures S1, S2, S3 and S4. All authors assisted in writing the manuscript.

\newpage
\section*{Supplementary information.}

\subsection*{Normalization and Cross-Talk}

Since the encoding technique employed for creating these holograms requires that the amplitude of the field be normalized to unity, energy conservation is violated in the generated modes. To compensate for this power scaling, a correction parameter is introduced for each transmission function. The correction parameter, $\alpha_{n}$, is calculated as the ratio between the encoded optical field $\tilde{\Psi}_n(\textbf{r})$ and the mode field $\Psi_{n}(\textbf{r})$, where $\alpha_{n} = \max\lbrace|\Psi_{n}(\textbf{r})|\rbrace^{-1}$ and $\alpha_{n} \in \Re^{+}$. In the case of the measurement procedure (i.e. performing the modal decomposition) the inner product can now be expressed as  

\begin{eqnarray}
\langle \tilde{\Psi}_n \vert \tilde{\Psi}_m \rangle = \langle \alpha_n \Psi_n \vert \alpha_m \Psi_m \rangle = \alpha_n \alpha_m \delta_{nm},
\label{eq:norm1}
\end{eqnarray}

where the correction coefficients are determined as a special case of equation (\ref{eq:norm1})

\begin{eqnarray}
\alpha_n^2 = \langle \tilde{\Psi}_n \vert \tilde{\Psi}_n \rangle
\label{eq:norm2}
\end{eqnarray}

From the detection signal (i.e. the measured on-axis intensity, $I_{n}(\textbf{r})$), the correction coefficients may be applied through the following relationship to normalise the signal to unit power:

\begin{eqnarray}
I_n(\textbf{r}) = \dfrac{\tilde{I}_n(\textbf{r})}{\alpha_n^2}
\label{eq:norm3}
\end{eqnarray}

\begin{figure}[h!]
\centering
\includegraphics[width = .5\textwidth]{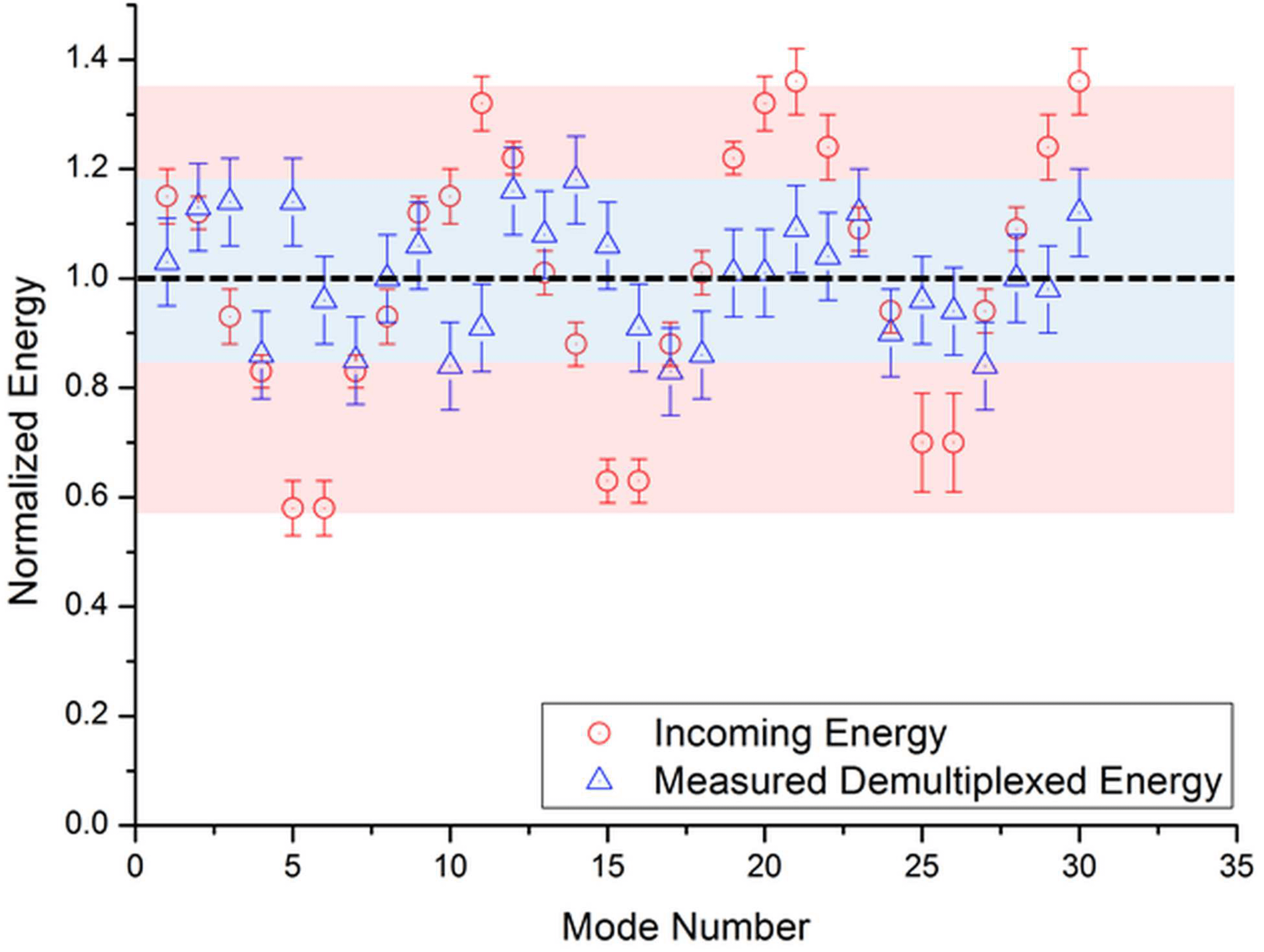}
\caption{{\bf Normalization of detected signals.} 
Plot of the the ratio of the energy in the generated modes with the energy in the detected signals as a function of the selected LG modes. The blue (red) data points contain the normalised (unnormalised) measured signals.}
\label{normalization}
\end{figure}

The normalization is illustrated in Fig.~\ref{normalization} where the ratio between the energy of the generated mode and the energy of the demultiplexed signal is plotted for the various LG modes being used. The red data points contain the unnormalised measured signals illustrating a wider spread from unity (marked by the dotted line), while the blue data points contain the normalised signals which have a narrower spread.    

\begin{figure}[h!]
\centering
\includegraphics[width = .5\textwidth]{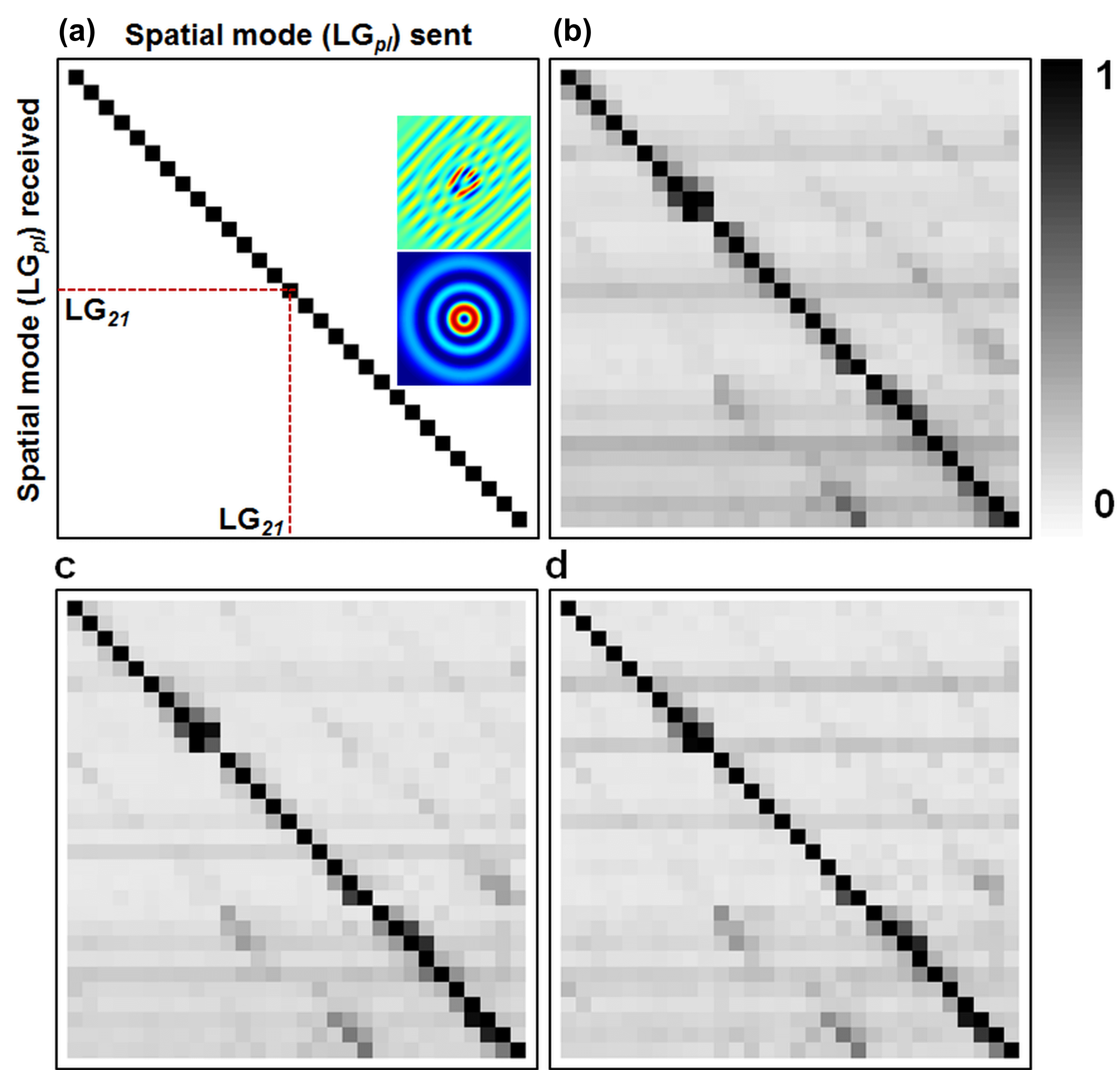}
\caption{{\bf Effect of aperture size in the detection plane.} Plot of the relative fractions of the intensity at each demultiplexed detector position for the selected LG modes. (a) The theoretical prediction and (b) - (d) measured cross-talk for aperture sizes of $24~\mu$m, $9.6~\mu$m and $4.8~\mu$m, respectively.}
\label{cross-talk}
\end{figure}

The effect of the aperture size in the detection plane when performing a modal decomposition on the incoming modes is also investigated. The selected LG modes (used as information carriers) were generated on SLM-1 and decomposed via an inner-product measurement at SLM-2 ( as depicted in Fig.~2). The measurement results are presented in Fig.~\ref{cross-talk} which illustrate the (a) expected and measured cross-talk between neighbouring modes for a detection aperture size of (b) $24~\mu$m (c) $9.6~\mu$m and (d) $4.8~\mu$m. In all cases [(b) - (d)] the strong diagonal and weak off-diagonal terms imply a highly accurate and precise measurement system which is unaffected by the size of the detection aperture. 

\subsection*{Gray-Scale and Colour}
\begin{figure}[h!]
\centering
\includegraphics[width = .5\textwidth]{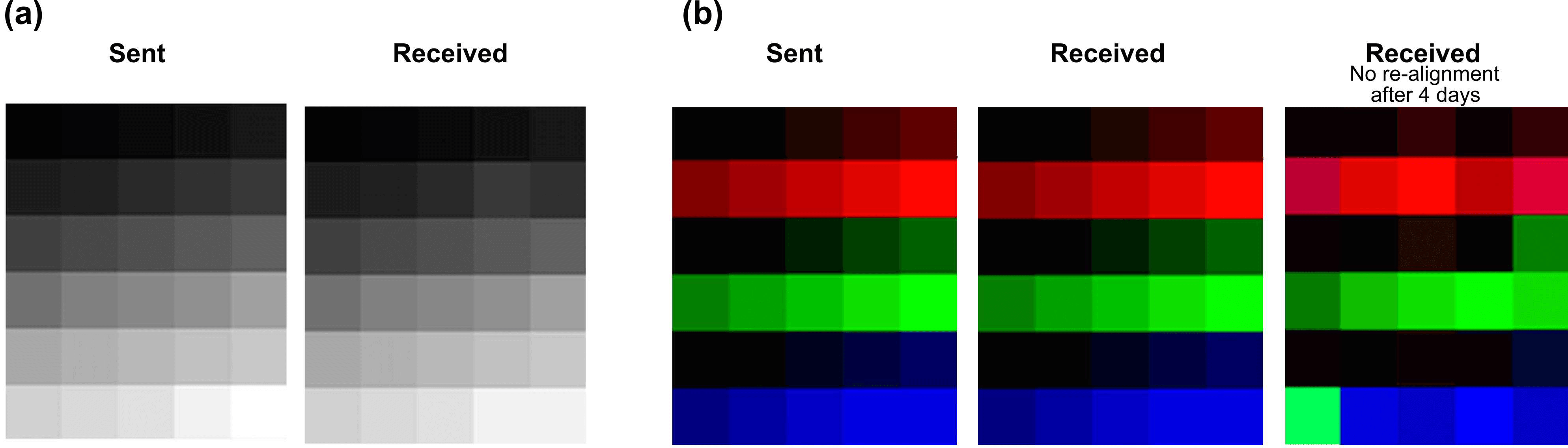}
\caption{{\bf Gray-scale test.} (a) The sent and received test image for verifying the success of the gray-scale encoding protocol. (b) The sent and received test image for verifying the success of the RGB encoding protocol.}
\label{gray}
\end{figure}

The concept of assigning colour values present in a 2D image [as depicted in Fig.~5 (a) and (b)] was initially tested with a simple image such as those in Figs \ref{gray} (a) and (b). This test required ensuring that the detector positions were aligned correctly with the on-axis demultiplexed signals. Incorrect alignment would result in incorrect colour values being decoded. The initial test images contain either 30 different gray-levels [\ref{gray} (a)] or RGB-values [\ref{gray} (b)]. In both cases it is  evident that the reconstructed images are in very good agreement with the sent images illustrating the correct alignment of the demultiplex signals with the static detector positions. The third image of Fig.~\ref{gray} (b) contains a reconstructed image acquired four days after the system was initially aligned, illustrating its robustness and versatility. 
\subsection*{8-Bit Encoding}
\begin{figure}[h]
\centering
\includegraphics[width = 0.45\textwidth]{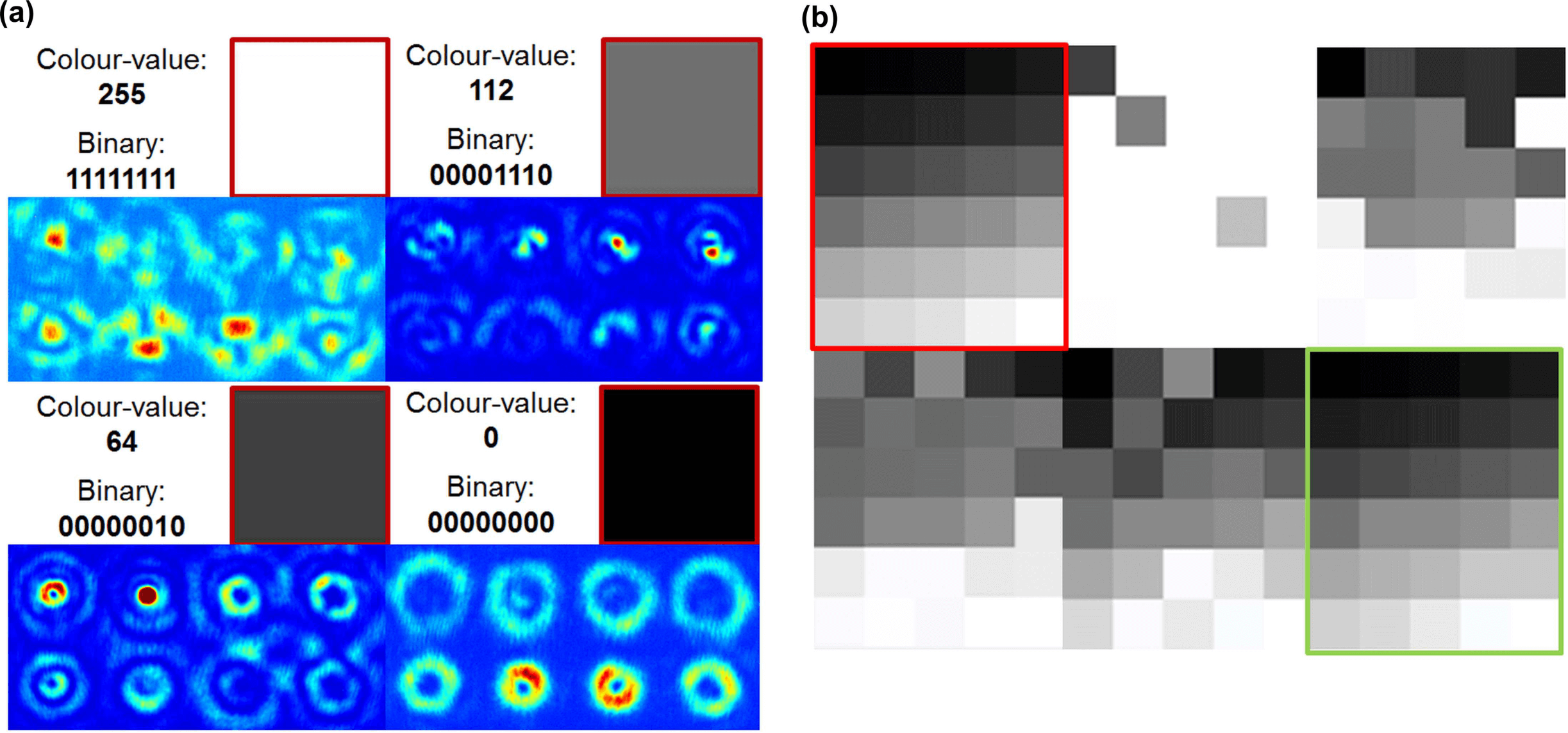}
\caption{{\bf Threshold test for 8-bit encoding scheme.} (a) Four selected gray-scales with their corresponding binary values and CCD images depicting the signals recorded at the detection plane. (b) The sent (red) and reconstructed gray-scale images for varying threshold values. Green marks successful reconstruction.}
\label{binary}
\end{figure}

In the gray-scale and RGB schemes the user is only concerned with detecting 1 signal out of a possible 35. However, the 8-bit scheme involves detecting 256 unique combinations of either no signal incremented in unit steps up to all possible 8 signals. Since the on-axis intensity is higher for a single signal as opposed to all 8 signals being present [demonstrated in Fig.~\ref{binary} (a)], the user needs to carefully select the range of thresholds for the measured intensities. We investigate the impact the intensity threshold has on our encoding scheme and our findings are presented in Fig.~\ref{binary} (b). The red border marks the sent image and the green border the successfully reconstructed image obtained at a suitable threshold. The images in between denote the reconstructed images when the threshold was initially set too low illustrating either extreme or mild cross-talk with neighbouring gray-values. Most of the cross-talk occurs with the white colour-value because when the threshold is set too low, the detectors detect noise - often resulting in 8 signals being detected which is the trademark of the white colour-value [Fig.~\ref{binary}(a)].

\subsection*{Beam quality factor}
\begin{figure*}[h!]
\centering
\includegraphics[width = .8\textwidth]{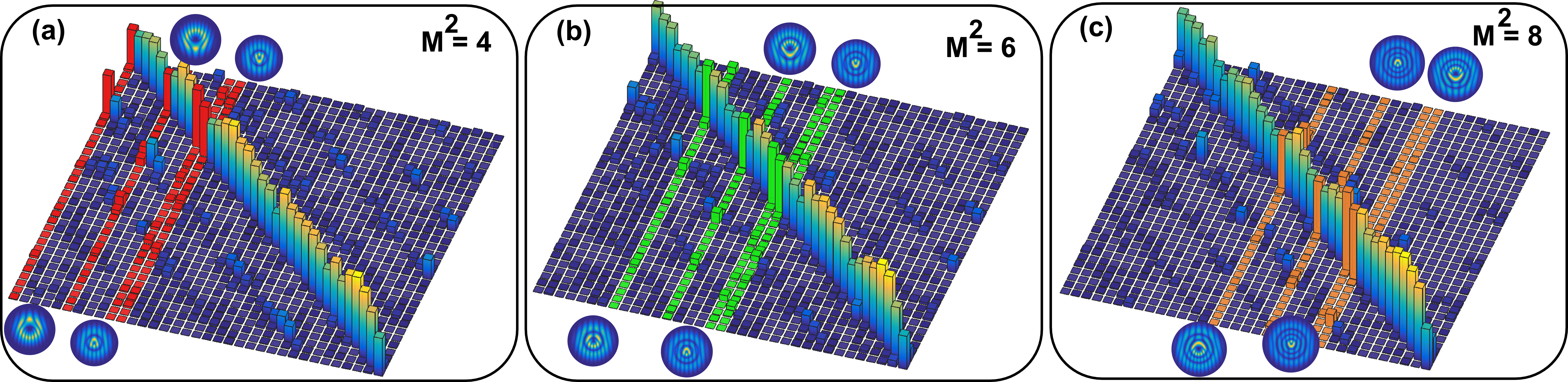}
\caption{{\bf Beam quality factor.} Cross-talk table for each wavelength showing the modes that shares the same beam quality factor. (a) $\lambda_1$ and $M^2=4$, (b) $\lambda_2$ and $M^2=6$ and (c) $\lambda_3$ and $M^2=8$.}
\label{BQF}
\end{figure*}

Through the beam quality factor, $M^{2}= 2p+|\ell|+ 1$ we can identify the LG modes that will propagate in an identical manner.  Figure \ref{BQF} shows the cross-talk table for each of the wavelengths we used in our experiment, in each table, we highlight some examples of the beams characterized by the same $M^2$ factor.  Fig.~\ref{BQF} (a) corresponds to $\lambda_{1}=457$ nm, here the lines in red show four modes $\mathrm{LG}_{0-3}$, $\mathrm{LG}_{11}$, $\mathrm{LG}_{03}$ and $\mathrm{LG}_{1-1}$  (see insets) that share the mode index, $M^2=4$. In Fig.~\ref{BQF} (b) we highlight in green the modes corresponding to $\lambda_{2}=488$ and $M^2=6$ which are $\mathrm{LG}_{1-3}$, $\mathrm{LG}_{2-1}$, $\mathrm{LG}_{13}$, and $\mathrm{LG}_{21}$. As a final example, Fig.~\ref{BQF} (c) shows the cross-talk table for $\lambda_{3}$=514 nm, highlighted in orange the modes with $M^2=8$. This is, $\mathrm{LG}_{3-1}$, $\mathrm{LG}_{23}$, $\mathrm{LG}_{31}$ and $\mathrm{LG}_{2-3}$.

\end{document}